\documentstyle[aps,epsfig]{revtex}
\newcommand{\be}{\begin{equation}}
\newcommand{\ee}{\end{equation}}
\newcommand{\bea}{\begin{eqnarray}}
\newcommand{\eea}{\end{eqnarray}}

\begin{document}
{\Large
\begin{center}
\bf{Combined effects of nuclear Coulomb field,  
    radial flow, and opaqueness on two-pion correlations}
\end{center}
}
\begin{center}
H. W. Barz\\
     Institut f\"ur Kern- und Hadronenphysik,\\
     Forschungszentrum  Rossendorf,\\
     Pf 510119,\\
      01314 Dresden, Germany\\

%\today\\
\end{center}
\vspace*{20mm}
 {\bf Abstract:}\\
Correlations of two like charged pions emitted from 
a hot and charged spherically expanding nuclear system
are investigated.
The motion of the pions is described quantum mechanically using 
the Klein-Gordon
equation which includes Coulomb field and pion absorption.
Flow modifies the radial distribution of the
source function and rescales the pion wave functions.
The radii extracted from the correlation functions 
are calculated in sideward  and outward direction
as a function of the pair momentum.
Comparison with recent measurements 
at SIS and AGS energies is made.

\vspace*{10mm}

\newpage
%%%%%%%%%%%%%%%%%%%%%%%%%%%%%%%%%%%%%%%%%%%%%%%%%%%%%%%%%%%%%%%%%%%%%%%%
\section{Introduction}

Measurements  of correlations  of pions and kaons as a function of their
relative momenta are used
to determine the source radius in heavy  ion reactions 
at intermediate
and relativistic energies.
On the basis of the Hanbury-Brown and Twiss
effect(see \cite{Kaufmann,bauer}) the correlation function
is related to the source radius via $R_0=\hbar c/\Delta q$,
where $\Delta q$ is the width of the correlation function. 
However it is well known
\cite{flows,hermbert} that the correlation is a complicated
function not only of the relative momentum but also 
of the total momentum of the pion pair. The correlation 
is determined by the size of the region from which
pions are emitted with roughly the same momenta. 
This has the  
consequence that for collectively streaming matter this 
region is smaller than the total source 
due to the  strong correlation between
the momenta and the emission points of the particles.
A further modification of the apparent source size
arises if pions are often rescattered within the source.
Thus, for pions the source is opaque \cite{Heiselberg},
and their emission points lie within a thin surface layer.

On the other hand the central Coulomb force 
changes the momenta of the particles while moving towards the 
detector. This latter effect was
investigated in refs. \cite{barzhbt,baymhbt}.  
Pions with small momenta are mostly influenced by the Coulomb forces 
which act quite differently on positively and negatively charged  pions.
Indeed quite different radii have been observed recently in collisions of 
Au on Au at bombarding energies of 1 GeV \cite{PelteHBT} 
and 11  GeV per nucleon \cite {barette} at 
the SIS and AGS accelerators, respectively.

The aim of this work is to study the combined effects
of central charge, opaqueness and flow on the extracted  radii. 
As a model we consider the emission of pions from a hot spherical 
nucleonic system the expansion of which can described by
radial hydrodynamical flow.  To ease the 
necessary numerical calculations we use a spherical spatial
distribution while for the momentum distribution a relativistic
Boltzmann distribution is used. 
Starting from a covariant formalism for two particle emission
(Sect.~II) we derive the source function in Sect.~III and obtain
a concise expression for the matrix element for the emission
of two particles. 
Once we have obtained the matrix element the standard
technique is applied to calculate the correlation function. 
Coulomb field and opaqueness are included
via mean fields in calculating the distorted waves 
for the pions. Numerical results are finally discussed in Sect.~VI.

\section {Basic equations}

Here, we briefly review the  formulation of the Hanbury-Brown and Twiss
effect which allows to incorporate the mean fields between
the source and the two emitted mesons. At asymptotic distance the 
two mesons move with momenta $\bf p$ and $\bf p' $. Each 
meson is described by a wave
function $ \psi_p(x) $ which satisfies an equation of motion
which contains the mean field, and $p=(\omega,\bf p)$ and $x=(t,{\bf r})$ 
denote \footnote{ The convention
$\hbar = c = k_{Boltzmann} =1$ is used.} the four-momentum and the 
position in time and space, respectively.  
It is customary to 
introduce a source term  $J(x)$ from which the wave function
$\psi_p$ is generated. With these definitions one can 
express \cite{Kaufmann,pawel} 
the probability for the emission of two meson in terms of the
source operators  $J^+(x), J(x)$ as
\bea  \nonumber
\omega \omega' \frac{dN}{d{\bf p} d{\bf p'}}&\, \sim \,&
 \int d^4x_1 d^4x_2 d^4x_3 d^4x_4 \times \\
&&\psi^*_p(x_1) \psi^*_{p'}(x_2)
                          \psi_{p'}(x_3) \psi_p(x_4)
             \langle J^+(x_3)  J^+(x_4)  J(x_1)  J(x_2) \rangle.
\label{dn1}
\eea

The basic assumption for applying interferometry to nuclei
is the chaoticity of the source i.e. the absence of initial correlations
between the two emitted pions except those correlations coming
from the Bose-Einstein statistics. Thus, one writes
\bea  \nonumber
&& \langle J^+(x_3)  J^+(x_4)  J(x_1)  J(x_2) \rangle  =\\
&& \langle J^+(x_4)  J(x_1) \rangle  \langle J^+(x_3)  J(x_2) \rangle +
 \langle J^+(x_3)  J(x_1) \rangle  \langle J^+(x_4)  J(x_2) \rangle.
\eea
This allows to express Eq.(\ref{dn1}) as products of 
density matrices
\be
\omega \omega' \frac{dN}{d{\bf p} d{\bf p'}} \, \sim \, 
\rho(p,p) \rho(p',p') + \mid \rho(p,p') \mid ^2
\label{dn2}
\ee
with 
\be
 \rho(p,p') =  \int \int  d^4x \, d^4x' \, \psi^*_p(x)  \psi_{p'}(x')
\langle J^+(x')  J(x) \rangle\;
\label{defmatrix}
\ee
which contains the one-body source term. This term 
can be connected to a classical source function $S(p,x)$ via
a  Wigner transformation 
\be
\langle J^+(x')  J(x) \rangle =
\int \frac{d^4k}{(2\pi)^4} e^{-ik(x-x')} S(k,\frac{x+x'}{2})\;.
\label{wigner}
\ee
The source function  $S(k,x)$ describes the creation 
of a meson at space-time $x$ with four momentum $k$.

The problem is further essentially reduced by assuming that the 
interactions do  not depend on time. 
This assumption allows the use of stationary solutions for
$\psi_p(x) = exp(-i\omega t)\,\psi_{\bf p}({\bf r})$ with 
$\omega = p^0  = \sqrt{m^2+{\bf p}^2}$.
Inserting the definition(\ref{wigner}) into Eq.(\ref{defmatrix}) and
integrating over the variable $t-t'$ we obtain
\bea   \nonumber
 \rho(p,p') &=& \int  \int d{\bf r}\, d{\bf r}'\, 
   \psi^*_{\bf p}({\bf r})
\psi_{\bf p'}({\bf r}') \\
&& \times \int \frac {d{\bf k}}{(2\pi)^3}\, e^{i{\bf k}({\bf r}-{\bf r}')}
  \int dt\;    e^{i(\omega-\omega ')t}
      \; S[(\frac{\omega+\omega'}{2},{\bf k}),(t,\frac{\bf r + \bf r'}{2})].
\label{matrix}
\eea
In the interference term for $p \ne p'$ the energy of the pion in the source 
function is fixed to the mean energy of the observed pions.  
In the case of a thermal momentum distribution 
without any correlation between space and momentum the integration
over ${\bf k}$ introduces correlations between ${\bf r}$  and ${\bf r'}$
in the order of the thermal wave length $\sqrt{2\pi/(mT)}$.
If plane waves are used the integration over the difference ${\bf r}
-{\bf r}'$ can be carried out which fixes the ${\bf k}$ momentum
to half the pair momentum leading to the standard form \cite{flows}
of the matrix element.

Now, the correlation function $C_2$ is defined as 
the two-particle emission
function normalized to the product of the one-particle 
emission functions which is given by the first part 
in Eq.(\ref{dn2}). It is convenient to introduce 
the average pair momentum ${\bf K}$  and the 
relative momentum ${\bf q}$  via
\be
{\bf K} = \frac{1}{2} ( \bf p + \bf p' ),\quad 
  {\bf q} =  \bf p - \bf p' \;.      \label{defKq}
\ee
Then, the correlation function reads
\begin{equation}
C_2({\bf{K}},{\bf{q}})\, = \,
  1 + \, \frac{ |\rho(K-\frac{q}{2}, K+\frac{q}{2})|^2}
          {\rho(K-\frac{q}{2}, K-\frac{q}{2})\,
          \rho(K+\frac{q}{2}, K+\frac{q}{2}) }\;.  \label{defHBT}
\end{equation}

For completeness we mention that the final state interaction 
\cite{prattcoul,anchiskinfinal} 
between the two outgoing mesons has not been considered. 
Inclusion of this interaction would need replacing the product 
of the two outgoing waves $\psi_{\bf p}(x) \psi_{\bf p'}(x')$ 
in Eq.(\ref{dn1}) by a 
correlated wave function $\Psi(x,x')$. Including the interaction 
with the source the function $\Psi(x,x')$ is, however, the complicated
solution  of a genuine three body problem.
As a first approximation one could assume that the source interacts 
mainly with the center-of-mass
of the pair while the final state
interaction is a function of the relative distance
between the outgoing particles only. 
As a consequence the function $\Psi$ could be split up
into a product leading to the standard expression 
\be
C_2({\bf{K}},{\bf{q}})_{final state} = P(q)\; C_2({\bf{K}},{\bf{q}})\;,
\label{fsi}
\ee
where the factor $ P(q)$ is in the simplest case 
the Coulomb penetrability. 
In the case of
Coulomb interaction such an approximation means that 
the quadrupole and higher momenta which influence the relative motion
of the pair are neglected. 
Alternatively, one can try to factorize the wave function $\Psi$
as it has been done in calculating proton-proton
correlations \cite{Erazmus}.

\section{Hydrodynamical picture}

Now we restrict our selves to a simple situation. We consider a hot 
spherical source from which pions are emitted.
Since the source expands the problem is not stationary. However,
we circumvent this difficulty by considering only  mesons which move 
with sufficiently large  velocities. Those
mesons are essentially outside the source
and always feel a time independent Coulomb
potential. Having in mind a collision of Au on Au nuclei at a
few $A\cdot$GeV generating a source of a temperature of about 100 MeV,
the thermal velocity of the protons
in the source is about 0.4 c embedded in a flow field of an average
velocity of about \mbox{0.3 c.}
Thus, the model may be applicable
for pion velocities larger than 0.7 c corresponding to
pion momenta above 100 MeV/c or kaon momenta above 300 MeV/c.
However, if the particles are released in the center of the source
they stay a while within the decreasing part of the Coulomb potential.
Within our model the only way to treat this problem is 
diminishing the central charge in a heuristic manner.
Such modifications of the effective charge have been observed 
\cite{Wagner} in
describing the $\pi^-/\pi^+$ ratio.

For ultrarelativistic energies the system 
does not expand spherically but is preferentially stretched in longitudinal 
direction. Since our treatment neglects
this dimension we do not consider the longitudinal correlation function.
Due to the longitudinal expansion the Coulomb force decreases with
time. As an approximate measure of the Coulomb action one can replace
the central charge $Z$ with twice the rapidity density 
$2(dN^+/dy - dN^-/dy)$
of the net charge as was shown in ref. \cite{BaBoGaHeII}.

In the hydrodynamical approach 
the momentum distribution is defined by a local temperature $T$ and 
a velocity field given by the four-vector $u^\mu$. 
Here we use typical parameters for 
a heavy ion reaction:
constant temperature $T$, radial mean velocity $\langle \beta \rangle$
and radial size $R_0$.  
Thus, we start with the source function 
\be
S(k,x)  =  \frac{1}{4\pi^2R_0^3\tau}
   e^{-\frac{ku}{T}-\frac{{\bf r}^2}{2R_0^2}\;
          -\frac{t^2}{2\tau^2}}
\label{sourcebasic}
\ee
The thermal distribution of the momenta within a 
fluid cell is coupled via $ ku = k^0u^0-{\bf ku}$ to the
four-velocity  $u^\mu$ of the cell. For large pion density 
one should replace the J\"uttner or relativistic Boltzmann distribution
with a 
Bose distribution. A spherically expanding system can described by
the flow velocity field
\be 
   u^\mu = (u^0,{\bf u}),\quad   
u^0  = \sqrt{1+{\bf u}^2}, \quad  
{\bf u} =\beta_0  {\bf r}/R_0,  \label{defu}
\ee 
where the four-velocity scales with the distance from the center.              
The parameter $\beta_0 $ is 
related to the mean flow velocity \mbox{$\langle \beta \rangle
= \langle |{\bf u}|/u^0 \rangle$} 
averaged  over the density (\ref{sourcebasic}) which 
leads to $\langle \beta \rangle = \sqrt{8/\pi}\beta_0 $  in
the limit of small $\beta_0 $.
Since $\beta_0$ characterizes a four-velocity it does not have
an upper bound while in contrary 
$\langle \beta \rangle$ never exceeds the velocity of light.
In Eq.(\ref{sourcebasic}) the pions are radiated off during
the emission time $\tau$.

Inserting the source function (\ref{sourcebasic}) into 
Eq.(\ref{matrix}) and integrating over the time we obtain 
\bea
\nonumber
\rho(p,p') \,&=& \,  \frac{1}{(2\pi)^{9/2}R_0^3} \;
            e^{-(\omega-\omega')^2\tau^2/2}
           \int d{\bf{r}} \, d{\bf r}' \,
              \psi^{*}_{{\bf{p}}}({\bf r })
        \psi_{{\bf{p}}'}({\bf r}')\\ 
  && \times \int  d{\bf k}    e^{-i{\bf k}({\bf r} - {\bf r'})}
       e^{{\bf k}{\bf u}/T}  e^{-({\bf r}+{\bf r'})^2/(8R_0^2)
   -(\omega+\omega')\sqrt{1+{\bf u}^2}/(2T)}\;,
\eea
where the velocity field ${\bf u}$ is proportional to 
$\frac{{\bf r} + {\bf r'}}{2}$. In the range of the convergence
of this integral the integration over ${\bf k}$ leads to a 
$\delta$ function
$ \delta({\bf r}- {\bf r}' -i{\bf u}/T)$ which
allows carrying out the integration over $ (\bf r -  r')$.
Thus, the final expression reads
\bea
\nonumber 
\rho(p,p') &=&    \frac{1}{(2\pi)^{3/2}R_0^3}\; 
        e^{-(\omega-\omega')^2\tau^2/2} \int d{\bf r} \, 
[\psi_{\bf p}(\alpha {\bf r} )]^*\, 
\psi_{\bf p'}(\alpha{\bf r})\\
&&\times  e^{- {\bf r}^2/(2R_0^2)-
(\omega+\omega')\sqrt{1+(\beta_0/R_0 \bf r)^2}/(2T)}\;.
\label{rhofinal}
\eea
The factor  $\alpha= 1-i\beta_0/(2TR_0) $ rotates the integration path
for the distorted wave functions $\psi_{\bf p}$ into the complex plane.

Using Eq.(\ref{rhofinal}) we obtain the correlation function
from Eq.(\ref{defHBT}). Since the system is spherically symmetric
we investigate the correlation as a function
of the the momentum-difference vector ${\bf{q}}_{out}$ pointing 
in the direction
of the average pair momentum ${\bf K}$ and  ${\bf{q}}_{side}$
being orthogonal to ${\bf{K}}$. 
It turned out that the correlation function $C_2$ can well be 
approximated by 
\begin{equation}
C_2= 1\;+\;\exp{(-{\bf  q }_{side}^2R_{side}^2
-{\bf q }_{out}^2R_{out}^2)}, 
\label{corrdef}
\end{equation}
where $R_{side}$ and $R_{out}$  characterize the extension of the source 
in and perpendicular to the direction of the pair momentum ${\bf K}$.

\section{Case of Zero Charge}

Before we turn to the full problem let us discuss the 
case of a negligible potential $U$. The distorted waves
in Eq.~(\ref{rhofinal}) can be replaced by plane waves
\be
       [\psi_{\bf p}(\alpha {\bf r})]^*
\psi_{\bf p'}(\alpha {\bf r}) =
       \exp{[-i {\it Re} (\alpha) {\bf q \bf r} +   
     2 {\it Im} (\alpha) \bf K \bf r]}
\label{planewave}
\ee
using the definitions (\ref{defKq}). To discuss the main effect
it  is instructive to find an analytical approximation. 
Expanding  the integrand in Eq. (\ref{rhofinal}) 
for small flow velocities up to order $\beta_0^2 $ 
and using $ \omega - \omega ' = {\bf K}{\bf q}/\tilde m $
for small momenta ${\bf q}$ one 
obtains from Eq. (\ref{defHBT}) the result
\be
C_2 = 1\;+\; \exp{\bigg\{-q^2\;\Big[(|\alpha|^2\, 
           \frac{R_0^2}{1+\beta_0 ^2 \tilde m/T})
           +(\frac{\bf qK}{q \tilde m})^2 
            \tau^2\Big] \;\bigg\} }.
\label{planeres}
\ee
with $\tilde m = \sqrt{m^2+{\bf K}^2}$.

Comparing the last equation with Eq.(\ref{corrdef}) one identifies 
the first expression in the round  bracket with the sideward
radius while the whole square bracket is the outward radius
with $|{\bf K}|/\tilde m$ being the pair velocity.
 The value of $|\alpha|^2$ exceeds unity by at most a  few percent 
for realistic 
values of flow velocities. Eq. (\ref{planeres})
contains the well known result \cite{flows}
that the radial flow reduces 
the observed radius $R_{side}$ 
with  increasing pair momentum and decreasing temperature. 
From the approximate behavior of the radius 
$R_0/\sqrt{1+\beta_0^2\tilde{m}/T}$
one recognizes that the dependence on the pair momentum
is essentially
a relativistic effect that is caused by the change of the relativistic 
pion mass
$\tilde m$. Eq. (\ref{planeres}) also contains the fact
that the outward radius is larger for a finite pion emission
time $\tau$. However, effect of mean fields could violate this
statement as shown in the following.

\section{Numerical Treatment}

To incorporate the mean field we solve the Klein-Gordon equation
\begin{equation}
\big[-\frac{\partial^2}{\partial{\bf{r}}^2}-
(\sqrt{m^2+{\bf{p}}^2}-U)^2 + m^2 \big]  \;
\psi^{(-)}_{{\bf{p}}}({\bf{r}}) = 0.
\label{kleingordon}
\end{equation}
The boundary conditions are chosen such that $\psi$
behaves asymptotically like an outgoing wave in the direction of ${\bf p}$
with incoming
spherical waves. This is indicated by the upper index (-).
We mention that functions with outgoing spherical waves
can also be used to calculate  the matrix element (\ref{defmatrix})
applying the relation $\psi^{(-)}_{\bf p} =
\psi^{(+)*}_{-{\bf p}}$.

The potential $U$ 
\be
U= \pm Z \, \frac{e^2}{r}\,\Phi(\frac{r}{\sqrt{2}R_0}) + 
i \, \frac{p}{\omega} \, \frac{1}{2 \lambda} \;,
\label{pot}
\ee
contains the Coulomb potential of the Gaussian source 
(\ref{sourcebasic}) with charge number
$Z$, and the quantity $\Phi$ denotes the error function. 
Further we include the possibility that
the pions might be absorbed within the source. 
For this purpose we have also introduced an
imaginary part 
which  depends on the mean free path $\lambda $.
The positive imaginary part ensures
that the wave function $\psi^{(-)}_{\bf p}$
increases in the direction of the outgoing momentum ${\bf p}$.
The potential (\ref{pot}) is only a rudiment
of the standard pion potentials used in calculating
pion-nucleus scattering and is usually derived as part in a
Schr\"odinger equivalent equation, see e.g. refs. 
\cite{Kisslinger,Siciliano}. 
However, the simple form in Eq.(\ref{kleingordon})
suffices for the study
of the effect of opaqueness which arises from the absorption and reemission
of pions in the matter.

The solution is numerically obtained by expanding 
the distorted wave into partial waves $f_l$
\bea
\psi^{(-)}_{\bf p}({\bf r}) 
= \frac{4\pi}{pr} \;\sum_{l\,m} i^l e^{-i\sigma_l}
          f_l(r) Y^*_{lm}({\bf p})  Y_{lm}({\bf r}),
\label{series}
\eea
where the quantities $Y_{lm}$ denote the spherical harmonics 
and the symbols $\sigma_l$ are the
Coulomb scattering phases. 
In order to use the non-relativistic 
standard method known from optical model calculations 
the numerical 
integration was extended to large radii $R_{max}$ to render the term $U^2$ in
Eq. (\ref{kleingordon}) negligible. 
Once the correct radial function has been obtained it is analytically
continued by integrating the radial differential equation from 
$r=R_{max}$ to $r=\alpha R_{max}$. The value obtained at this point
is used to normalize the function  $\psi(\alpha r) $ obtained by
integrating the  radial equation  along the path $ \alpha r $.

\section { Results and Discussion}
We study the model for a situation which is typical 
for a collision of Au on Au nuclei at bombarding energies
of 1 GeV per nucleon. 
In nearly central collisions 
a system of charge $Z=120$ is formed which a temperature of
about $T=80$  MeV for pions and a flow velocity of 
about $\langle \beta \rangle = 0.32$ \cite{pelte0,muntz}.
We use a source radius of  $R_h=10$ fm which corresponds to
the parameter $R_0=R_h/\sqrt{5}$. 

The correlation function (\ref{defHBT}) has been calculated for various
pair momenta ${\bf K}$ as function of the relative momentum $q$.
In all cases the obtained correlation function has 
nearly a Gausian shape
and is fitted  to Eq.(\ref{corrdef}) in the region of $C_2=1.5$
to obtain the HBT radii $R_{out}$ and $R_{side}$.

In Fig.  1  the ratios of the fitted source radii to the true
radius are displayed 
as a function of the averaged pair momentum
$|{\bf K}|$ for two 
extreme values of the flow velocity  $\langle \beta \rangle = 0$
and  $\langle \beta \rangle = 0.5 $. Compared to $R_0$ the
observed radii $R_{side}$ are increased for negatively charged pion
and diminished for positively charged pions
in comparison to the true radius. An opposite but smaller effect is seen for
the radii $R_{out}$. These changes are significant
for pions with momenta below 300 MeV/c.
Results for pair momenta smaller than 100 MeV are not shown
since the stationary approach is not justified. Below the
Coulomb threshold the behavior of the radius changes
drastically, see ref. \cite{barzhbt} for details. 
For comparison we have
inserted into right hand panel of Fig. 1 
the sideward radius extracted for $\pi^0$ mesons.
This curve nearly averages the lines for the unlike charged pions.
Comparing the curves for the two flow velocities
one recognizes that the corrections arising from the Coulomb
field and the flow field add up nearly
independently.
Fig. 2 shows the ratio $R_{HBT}/R_0$ for a duration of 
the emission of $\tau = 4$ fm/c which leads to the expected 
increase of the outward radius.

Differences between extracted HBT radii
for positively and negatively pions have been measured
at AGS \cite{barette} in the projectile rapidity region.
A ratio of $R_{side}(\pi^-\pi^-)$ to $R_{side}(\pi^+\pi^+)$ of
$(5.6\pm0.7)$ fm / $(3.9\pm0.8)$ fm = $1.4\pm0.3$
has been found which agrees with our predictions.
The ratio of the outward
radii of $(5.8\pm0.5)$ fm / $(6.5\pm0.5)$ fm = $0.9  \pm 0.15$
is smaller than unity although the relatively large error
bars do not allow a definite comparison. 
These  measurements
also  qualitatively agree
with the ratio of $1.2 \pm 0.4 $ of the sideward radii observed
by Pelte at al.\cite{PelteHBT}. In those measurements the
same increase was found for the radius $R_{out}$ contradicting
our predictions.

Now we investigate the effect of opaqueness of the source.
In ref. \cite{Heiselberg} a drastic change 
of the the HBT radii  was predicted. 
Pions with
momenta around $k_0$ = 270 MeV/c have a large total cross section
with nucleons exciting  strongly the $\Delta$ resonance.
Therefore, those pions have their last interaction points
within a thin surface zone near the direction of their momenta. 
The thickness of this zone is determined by the mean free path
$\lambda$ of the pions.  The inverse path length is estimated to be
\be
\frac{1}{\lambda_{\pi^{\pm}}} = n\;
        (\sigma_n^{\pi^\pm} \frac{N}{A}   \;+\;
               \sigma_p^{\pi^\pm} \frac{Z}{A}) 
\label{path}
\ee
which is proportional to the baryonic  density $n$ and the 
total cross sections $\sigma_N^{\pi^\pm}$ of pion-nucleon
collisions averaged over the thermal motion of the nucleons.
Due to isospin 
coupling the cross sections $\sigma_n^{\pi^-} = \sigma_p^{\pi^+}$
are by a factor of three larger than the remaining two.
This creates an isospin asymmetry of the mean free path
in neutron rich matter.
The thermal motion widens the $\Delta$ resonance to $\Gamma = 240$ MeV
and reduces the
maximum cross section by about a factor of two resulting
in a value of 100 mb for $\sigma_n^{\pi^-}$. 
At normal nuclear matter density of $n_0$ = 0.16 fm$^{-3}$
one obtains for heavy nuclei like Au values of 
$\lambda^0_{\pi^+}$= 1.05 fm  and $\lambda^0_{\pi^-}$ = 0.85 fm. 
Similar values are known from BUU calculations \cite{wolf}.
Now we can simulate the opaque source by introducing
the momentum and density dependent 
mean free path $\lambda = \lambda^0 n_0/n (1+(2(p-k_0)/\Gamma)^2)$
into the potential (\ref{pot}).

Fig. 3 shows the effect of the opaqueness 
for the Gaussian density distribution (\ref{sourcebasic})
with $R_h$=10 fm  for 
positive and negative pions without considering the 
electric charge. The opaqueness increases the radius $R_{side}$ 
while the radius  $R_{out}$ decreases as a consequence 
of the relatively thin middle part of the half-moon shaped
source region \cite{Heiselberg}. 
The curves reflect the resonance shape of the absorption. 
An essential effect of the opaqueness is that the difference
$R_{out}-R_{side}$ could  become negative which may 
compensate the positive contribution from the emission time.
In the ultra-relativistic regime
one should also add the effect of the $\pi^-$ $\pi^+$ scattering
since the pion density is large and the cross section
could reach values up to 15 mb for pion momenta around
200 MeV/c.

A strong dependence
of the side-correlation on the pion-pair momentum
has been found for the collision of Au on Au at a bombarding
energy of 1.06 GeV per nucleon in ref. \cite{PelteHBT}. 
A comparison to that data give us a good opportunity to 
illustrate how the different effects discussed so far
change the true radius. 
We employ the parameters $T=80 MeV$ and $\langle \beta \rangle=$
0.35 as before, however we reduce the central charge to
$Z_{eff}= 60$ to diminish the Coulomb effect trying
to correct for the expansion during the pion emission.
Using a  hard sphere radius of $R_h=8$ fm and an emission time of 
$\tau=4 $ fm/c the obtained sideward and outward radii are shown together
with the measurements \cite{PelteHBT} in Fig. 4. 

We do not intend to fit the data to our parameters
since our simple model lacks essential features, especially 
the time evolution of the collision treated in recent dynamical models.
Fig. 4 shows however that essential deviations from the true
radius are to be expected and high precission measurements
are needed to gain insight in the dynamics of nuclear collisions. 
For negative pions it is clearly seen that the measured sideward radius 
depends stronger on the pair momentum than one would expect from
calculations using a fixed radius. This means that indeed  the
fast pions come from an earlier more compressed stage of the
matter while the low energy pions are emitted
from a zone with a larger size.

\section{Conclusions}

The  nuclear Coulomb field 
increases (decreases) the observed HBT radii
extracted from sideward correlations for negatively (positively)
charged pions.
The influence is opposite
and smaller for the outward correlation. This effect is the largest for
small momenta and is superimposed
on the overall reduction caused by the radial flow.
The opaqueness due to pion rescattering leads to 
a decrease of the outward radius and an increase of the 
sideward radius. The decrease could compensate the
general increase of the outward radius due to the duration 
of the emission process.

\section*{Acknowledgments}
Support by the German ministry BMBF under contract 06DR829 is
acknowledged.

\newpage
%%%%%%%%%%%%%%%%%%%%================================

\newpage
{\bf Figures }

\begin{figure}[t]
\begin{center}
\mbox{\psfig{figure=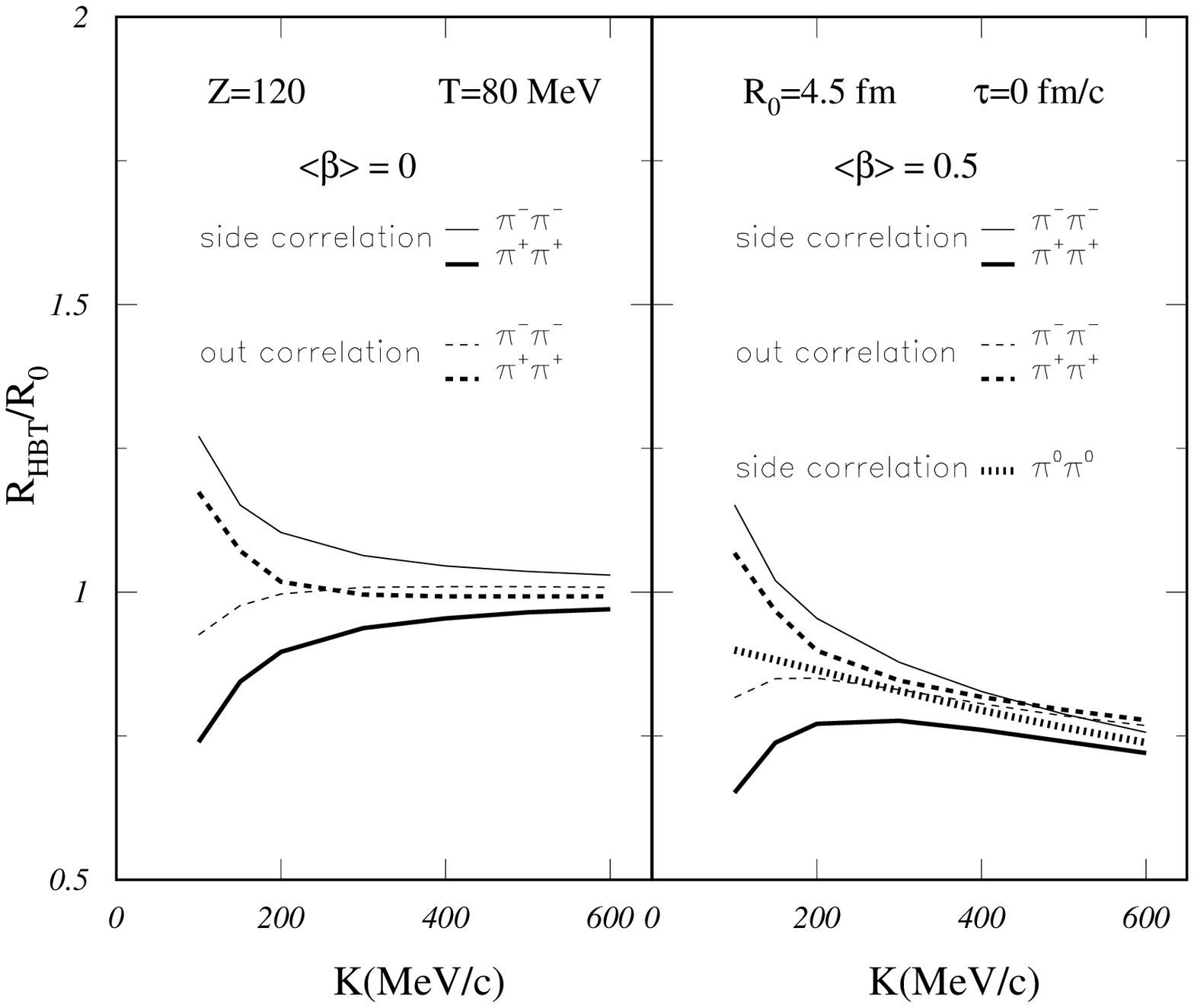,height=3.5in,width=110mm}}
\end{center}
\end{figure}
Fig. 1. Ratios of sideward and outward HBT radii to the true 
radius $R_0$ of a Gaussian shaped source 
as a function  of half the pair momentum $K$.
The ratios have been calculated without (left panel)
and with (right panel) radial flow of mean velocity 
$\langle \beta \rangle$. 

%\newpage
\begin{figure}[t]
\begin{center}
\mbox{\psfig{figure=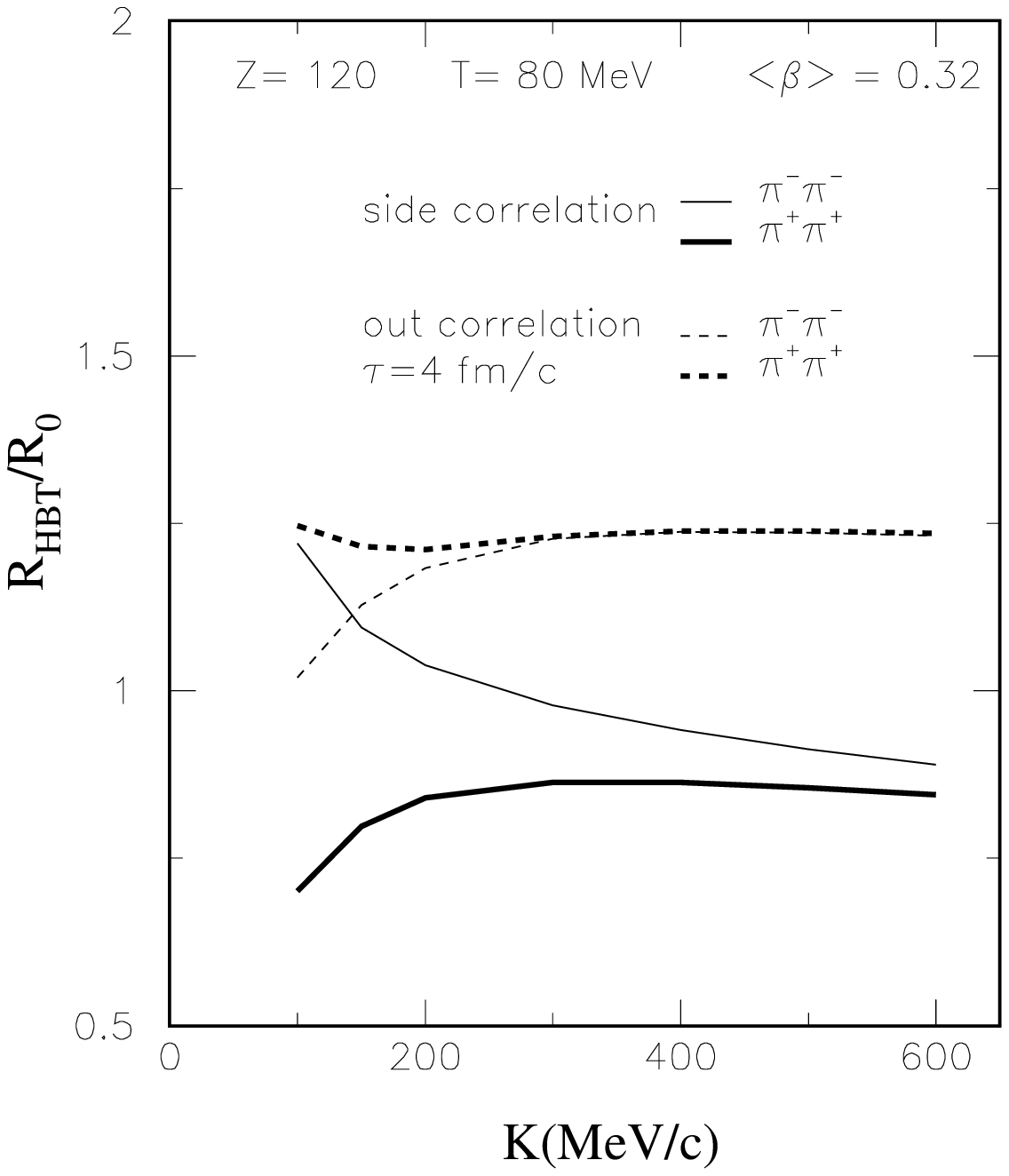,height=3.5in,width=80mm}}
\end{center}
\end{figure}
\vspace*{-20mm}
Fig. 2. Ratios of sideward and outward HBT radii to the
radius $R_0$ of a Gaussian shaped source with
radial flow for an emission time of 4 fm/c. 

\newpage
\begin{figure}[t]
\begin{center}
\mbox{\psfig{figure=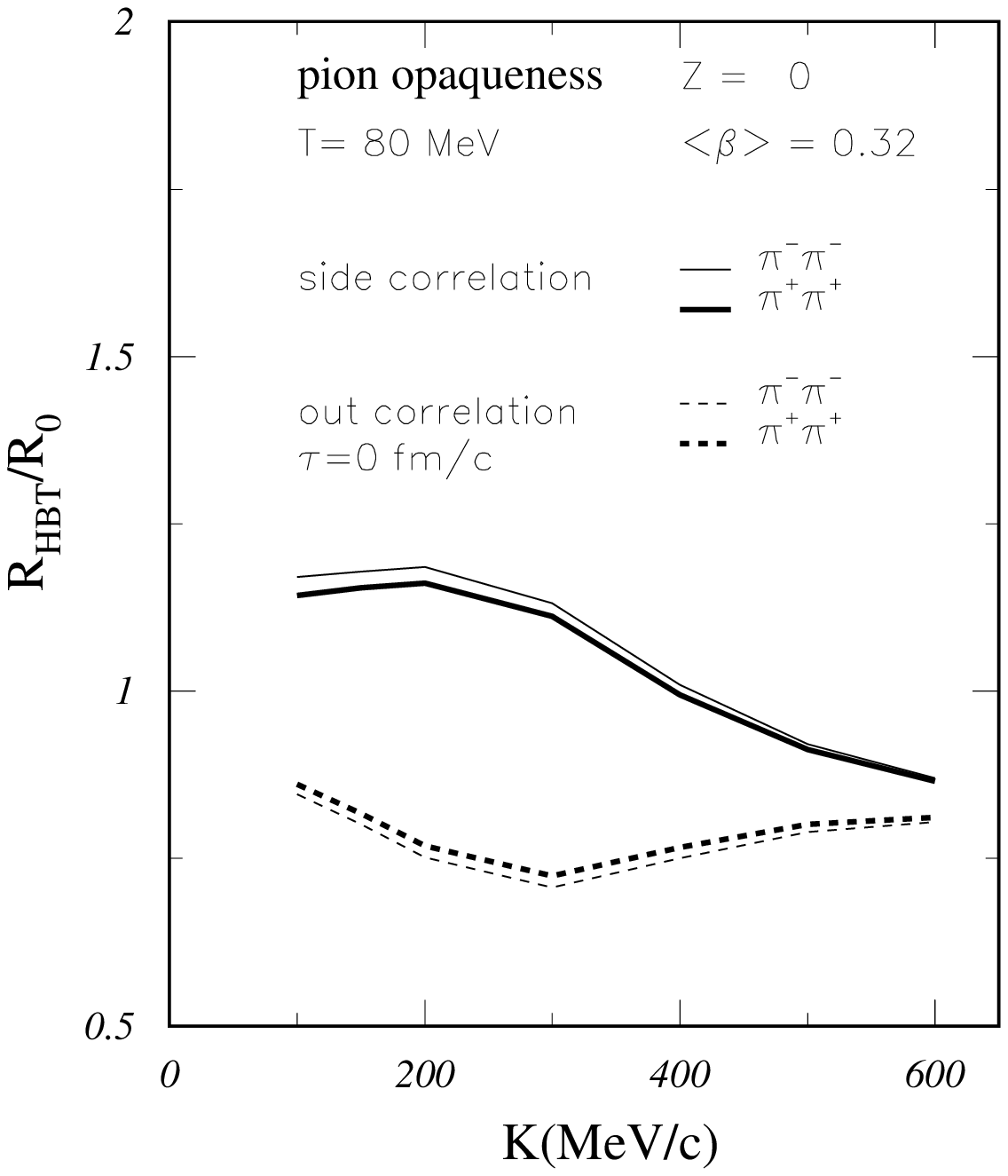,height=3.5in,width= 80mm}}
\end{center}
\end{figure}
\vspace*{-20mm}
Fig. 3. Ratios of sideward and outward HBT radii to the true
radius $R_0$ affected by flow and 
opaqueness of the source as a consequence of the small 
mean free path of pions within the source.

%\newpage
\begin{figure}[t]
\begin{center}
\mbox{\psfig{figure=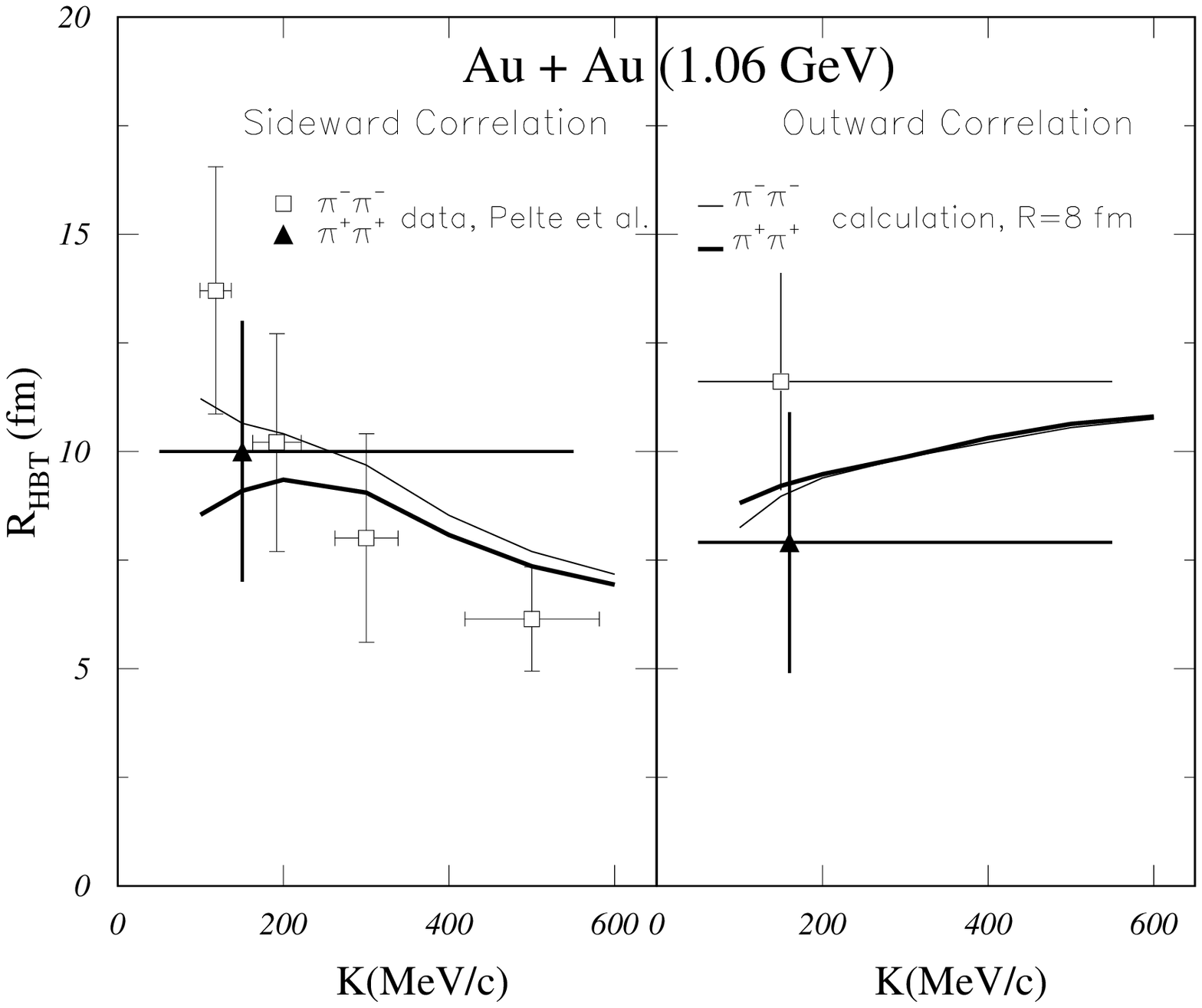,height=3.5in,width=110mm}}
\end{center}
\end{figure}
Fig. 4. Comparison of measured sideward and outward radii \cite{PelteHBT}
to calculated HBT radii affected by flow, central Coulomb field
and opaqueness as a function  of half the pair momentum. 
The large extension of some of the horizontal error bars 
indicates that the full 
range  of pair momenta has been used to extract the radius.
The calculation are carried out with a fixed source radius
of $R_h=$ 8 fm which cannot fully explain the momentum dependence 
of the measured sideward radius.

\end{document}